\documentclass[fleqn,10pt]{wlscirep}
\usepackage[utf8]{inputenc}
\usepackage[T1]{fontenc}
\usepackage{amsmath}
\usepackage{amssymb}
\usepackage{bm}
\usepackage{placeins}
\usepackage{xcolor}
\usepackage{lineno}
\usepackage[normalem]{ulem}
\hypersetup{hidelinks}


\title{DeepBD: A Grounded Agentic Workflow for Variant Prioritization and Diagnosis of Genetic Birth Defects}

\author[1,+]{Shiyu Li}
\author[1,+]{Ziqi Yan}
\author[1,+]{Zhihao Wu}
\author[1,+]{Jielong Lu}
\author[1]{Weiran Liao}
\author[1]{Jiajun Yu}
\author[1]{Genjie Li}
\author[1]{Zeyu Chu}
\author[1]{Jiajun Bu}
\author[1,*]{Haishuai Wang}
\affil[1]{College of Computer Science and Technology, Zhejiang University, Hangzhou, China}
\affil[+]{These authors contributed equally to this work.}
\affil[*]{Corresponding author: Haishuai Wang (haishuai.wang@zju.edu.cn).}

\begin{abstract}
Birth defects are a major cause of fetal loss, neonatal morbidity and long-term disability. In the subset with suspected genetic etiologies, exome and genome sequencing have moved many cases from variant detection to post-sequencing interpretation: clinicians must rank patient-specific candidate variants under incomplete fetal or infant phenotypes and heterogeneous evidence from population genetics, variant-effect prediction, gene-disease validity, phenotype ontologies, cellular and pathway context, protein structure and clinical literature. We present DeepBD, a grounded agentic workflow for variant prioritization and diagnostic interpretation of genetic birth defects. DeepBD organizes the workflow into LLM-assisted case structuring, a pretrained evidence engine, specialist evidence modules and a grounded diagnostic review layer. The evidence engine learns patient-specific variant scores from structured rule evidence, sequence and variant-effect representations and phenotype-conditioned biological context, whereas specialist modules and the agentic layer provide tool-based refinement, candidate-pool review and diagnosis-oriented synthesis from ranked candidates. Developed using an in-house fetal and infant cohort comprising 18,622 cases, DeepBD achieved Recall@1/3/5/10 of
0.658/0.882/0.912/0.929 on an internal held-out solved-case benchmark, outperforming standalone Exomiser, DeepRare and prompted LLM reranking baselines evaluated on Exomiser-derived top-20 candidate variants. Ablation and overlap analyses indicate that rule evidence, mechanism context and specialist refinement contribute complementary signal. These results support a grounded agentic approach in which data-driven evidence integration, tool-based refinement and LLM-assisted diagnostic review are assigned distinct roles in genetic birth defect interpretation and diagnosis, with quantitative evaluation
focused on retrospective variant-prioritization performance.
\end{abstract}

\begin{document}

\raggedbottom
\maketitle
\thispagestyle{empty}

\section{Introduction}

Birth defects, also referred to as congenital anomalies, congenital disorders or congenital malformations, comprise structural or functional abnormalities that arise during intrauterine life and may be recognized prenatally, at birth or during infancy\cite{who_congenital_2023,cdc_birth_defects_2026,nichd_congenital_2024}. They represent 
a substantial global health burden, affecting an estimated 6\% of babies 
worldwide and contributing considerably to neonatal and under-5 mortality \cite{who_congenital_2023,cdc_birth_defects_2026}. Birth defects have heterogeneous causes, including genetic, environmental, infectious and multifactorial factors. In this manuscript, we use genetic birth defects to denote birth defects or early-onset congenital presentations with suspected chromosomal or single-gene etiologies. For this clinically actionable subset, a molecular diagnosis can inform fetal prognosis, perinatal management, neonatal care, recurrence-risk counseling and reproductive decisions\cite{miller2010chromosomal_microarray,monaghan2020fetal,vora2023prenatal}. Large developmental-disorder studies have also shown that pathogenic de novo and rare coding variants make a substantial contribution to severe early-onset developmental phenotypes, many of which overlap the clinical space of birth defects\cite{ddd2015large_scale,mcrae2017dnm}.

Genomic testing has changed where the hardest part of diagnosis lies. Chromosomal microarray, exome sequencing and genome sequencing can identify candidate molecular causes after fetal structural anomalies or early infant presentations, and multiple studies have established the clinical value of sequencing in these settings\cite{lord2019page,petrovski2019fetal,fu2022prenatal,mellis2022prenatal,p3egs2023diagnostic}. The post-test workflow remains difficult. A typical case may contain thousands of variants before filtering, several plausible genes after annotation and at least one variant of uncertain significance. Prenatal and pediatric sequencing studies have repeatedly emphasized that yield depends not only on sequencing technology, but also on indication, phenotype quality, family structure, variant interpretation and reanalysis capacity\cite{monaghan2020fetal,vora2023prenatal,p3egs2023diagnostic}. The result is a post-sequencing bottleneck: the system must determine which variant is causal for the current fetus or infant, which evidence is reliable and how a ranked candidate should be converted into a clinically meaningful diagnostic hypothesis.

Genetic birth defects impose a concentrated version of this bottleneck. Fetal phenotypes are often observed through ultrasound, magnetic resonance imaging or other indirect measurements; the phenotype profile may be incomplete because many developmental features are not yet visible; and additional features may appear only after birth or later clinical follow-up. Human Phenotype Ontology (HPO) terms provide a computable language for phenotypic abnormalities, and phenotype-driven systems have become central to genomic diagnostics\cite{robinson2008hpo,kohler2021hpo}. However, interpretation of genetic birth defects also requires connecting early organ-system phenotypes to candidate genes through cellular, anatomical and pathway contexts. This is why phenotype resources, genotype-phenotype knowledge graphs and case-report search systems are useful components of the workflow\cite{shefchek2020monarch,fujiwara2018pubcasefinder}. A robust system must preserve the patient phenotype, variant evidence and biological context in a form that can be scored, audited and reviewed.

Computational variant-prioritization tools have addressed important parts of this problem. Phen-Gen, Exomiser, Xrare and LIRICAL integrate phenotype information with genetic evidence to prioritize genes, variants or candidate diseases\cite{javed2014phengen,smedley2015exomiser,li2019xrare,robinson2020lirical}. Variant-effect predictors and annotation resources such as SIFT, PolyPhen-2, MutationTaster, CADD, REVEL, PrimateAI, SpliceAI and AlphaMissense provide complementary evidence about molecular consequence\cite{ng2003sift,adzhubei2010polyphen,kircher2014cadd,schwarz2010mutationtaster,ioannidis2016revel,sundaram2018primateai,jaganathan2019spliceai,cheng2023alphamissense}. Clinical databases and curation frameworks including ClinVar, ClinGen, HGMD, OMIM and ACMG/AMP guidelines provide auditable human evidence and rules for variant interpretation\cite{richards2015acmg,strande2017clingen,landrum2016clinvar,stenson2020hgmd,amberger2019omim}. These systems and resources have made genomic interpretation more systematic. The remaining gap is how to organize heterogeneous signals for the birth-defect setting: standalone prioritizers, molecular predictors and clinical databases each provide useful evidence, but their outputs must still be weighted for a patient-specific fetal or infant phenotype, reconciled when incomplete or conflicting and translated into a reviewable diagnostic hypothesis.

Learning-based prioritization methods make the data requirement of this problem explicit. Xrare jointly models phenotype and genetic evidence, DeepPVP uses deep learning for phenotype-based causal-variant prioritization, MAVERICK applies deep structured learning to Mendelian variant prioritization and AI-MARRVEL uses a knowledge-driven random-forest system trained on millions of variants from diagnosed cases\cite{li2019xrare,boudellioua2019deeppvp,danzi2023maverick,mao2024aim}. SHEPHERD further shows that knowledge-guided learning can support phenotype-driven rare-disease diagnosis, while also illustrating how simulation and few-shot strategies are used when directly labeled patient data are scarce\cite{alsentzer2025shepherd}. These studies argue against a simple rule-versus-learning dichotomy. The more relevant question is where learning is supported by real case distributions and where external evidence should remain modular. For genetic birth defects, public resources contain rich disease, variant and ontology knowledge, and prenatal sequencing studies provide important cohorts and standards\cite{lord2019page,fu2022prenatal,mellis2022prenatal,p3egs2023diagnostic}. Yet the specific training substrate needed for a birth-defect prioritization engine--fetal or early-infant phenotypes, sequencing-derived competing variants from the same patient, clinically curated causal labels and realistic post-sequencing noise--is difficult to obtain at scale.

LLM-based and agentic medical AI systems have recently expanded the design space. Med-PaLM and AMIE showed that large language models can encode clinical knowledge, answer medical questions and support diagnostic dialogue under structured evaluation\cite{singhal2023medpalm,tu2025amie}. DeepRare demonstrated an agentic rare-disease system with traceable reasoning, and Hygieia and Berrylyzer explored related agentic designs for rare-disease and prenatal genetic diagnosis\cite{zhao2026deeprare,liu2026hygieia,meng2026berrylyzer}. These studies show the value of language models for clinical-language normalization, retrieval, tool coordination, reflection and human-readable synthesis. Genetic birth defect interpretation places a different load on such systems. The central decision is usually not open-ended diagnostic brainstorming, but a ranked comparison among patient-specific candidate variants under incomplete developmental phenotypes. General-purpose agent loops can retrieve and summarize evidence, but variant-level causal ranking requires a stable evidence substrate that can learn from prior fetal and infant cases, incorporate explicit clinical rules and treat external tools as calibrated evidence rather than unconstrained reasoning steps. This motivates a grounded agentic design in which evidence computation, specialist tool use and LLM-based review are assigned distinct roles.

DeepBD was designed around this evidence-allocation principle. The system first initializes a pretrained evidence engine on in-house fetal and infant variant data and then optimizes it for causal variant prioritization using curated solved cases. For each patient and each candidate variant, the evidence engine constructs a structured substrate from sequencing-derived variant features, explicit clinical rules and phenotype-conditioned biological context. Specialist modules then add selected external evidence signals around this substrate. A grounded diagnostic agent assists case structuring, organizes provenance-preserving evidence, coordinates refinement modules and supports top-k review and diagnosis-oriented synthesis from ranked candidates. This design uses agentic capabilities for curation, routing, review and evidence synthesis while placing variant-level scoring in a trainable model that can be benchmarked and ablated.

Here we describe the DeepBD workflow and evaluate it in an in-house fetal and infant cohort with sequencing and phenotype data. We focus on genetic birth defects as a fetal-to-early-infant diagnostic continuum, because prenatal findings and early postnatal phenotypes often belong to the same molecular diagnostic pathway. In an internal solved-case benchmark,  we show that DeepBD improves causal variant prioritization over established tool-based, LLM and agentic reranking baselines, recovers a complementary subset of cases and derives its performance from multiple evidence levels. The results support an evidence-grounded, agentic approach for post-sequencing interpretation and diagnosis of genetic birth defects.

\section{Results}

\subsection{Grounded agentic workflow}

\begin{figure}[!t]
\centering
\includegraphics[width=\linewidth]{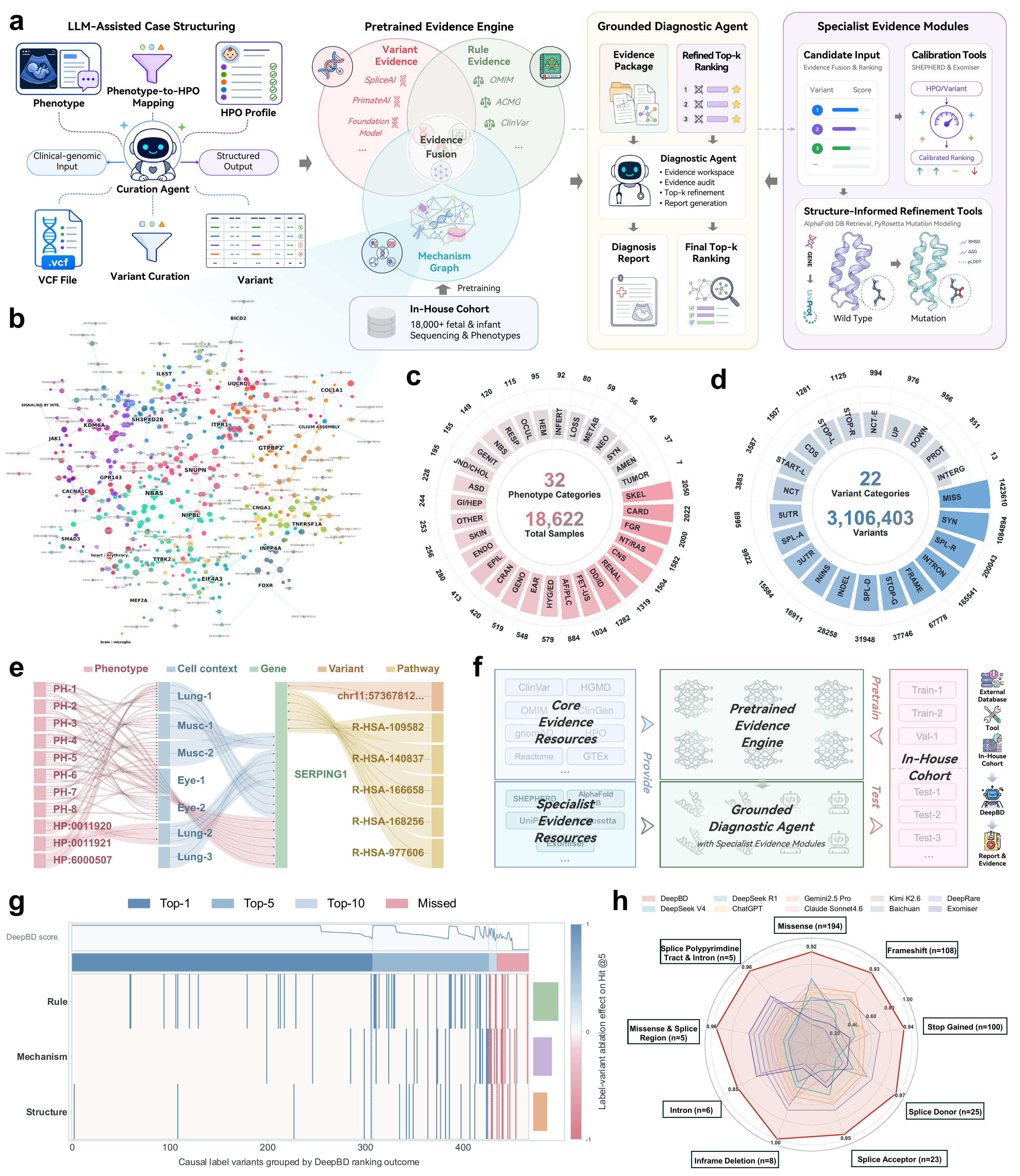}
\caption{\textbf{DeepBD workflow and cohort landscape.}
\textbf{a,} Workflow for genetic birth defect interpretation from clinical-genomic input to ranked variants and diagnosis-oriented evidence synthesis.
\textbf{b,} Biomedical graph context linking phenotype, cellular context, pathway, gene and variant evidence.
\textbf{c,d,} Phenotype and variant composition of the in-house fetal and infant cohort.
\textbf{e,} Example phenotype-cell-gene-variant-pathway flow for patient-specific biological context.
\textbf{f,} Evidence-resource and evaluation overview.
\textbf{g,} Distribution of curated positive variants across DeepBD ranking outcomes together with rule, mechanism and structure evidence patterns.
\textbf{h,} Variant-class-stratified Hit@5 comparison across DeepBD and baseline reranking systems.}
\label{fig:overview}
\end{figure}
\FloatBarrier

DeepBD takes as input a patient phenotype profile and a sequencing-derived candidate-variant table. The phenotype input can be structured HPO identifiers or free-text clinical descriptions that are mapped to HPO concepts for downstream computation\cite{robinson2008hpo,kohler2021hpo}. In a prenatal case, the phenotype may consist of ultrasound or imaging findings such as increased nuchal translucency, congenital heart defect, skeletal dysplasia, renal anomaly or fetal growth restriction. In an infant case, it may include postnatal examination, growth, neurodevelopmental or laboratory findings. The genomic input is a VCF- or TSV-derived table containing genomic coordinates, reference and alternate alleles, affected gene, consequence annotation, genotype, transcript or HGVS description and sequencing quality fields. DeepBD processes one patient case at a time, scores retained candidate variants within that case and returns a ranked table with evidence fields for review.

The workflow was designed around an evidence-allocation principle (Fig.~\ref{fig:overview}a). Genetic birth defect interpretation contains tasks with different computational character. Phenotype normalization, candidate-pool review and evidence synthesis are language-heavy and benefit from LLM-assisted structuring. Variant ranking is a high-dimensional, case-conditioned comparison problem and benefits from a trainable evidence engine. External algorithms such as SHEPHERD, Exomiser and structure modeling are useful specialist evidence sources, but their outputs need to be interpreted in the context of the patient. DeepBD therefore organizes the workflow into four interacting layers: LLM-assisted case structuring, a pretrained evidence engine, specialist evidence modules and a grounded diagnostic agent.

The case-structuring layer converts heterogeneous clinical-genomic input into a computable workspace. Clinical descriptions are mapped to HPO terms while preserving the original text, because fetal and infant phenotypes can be incomplete, imaging-derived or phrased differently across sites. Sequencing results are normalized into per-variant records with genomic coordinates, gene assignment, transcript-level consequence, zygosity and quality fields. This layer is deliberately light: it prepares structured inputs and preserves provenance, but it does not determine the causal variant.

The pretrained evidence engine is the central ranking component. For each candidate variant, it builds three core evidence representations. First, rule evidence encodes clinically interpretable signals such as allele frequency, consequence, dosage sensitivity, gene-disease validity and curated variant evidence\cite{karczewski2020gnomad,strande2017clingen,landrum2016clinvar,stenson2020hgmd,richards2015acmg}. Second, variant-intrinsic evidence represents sequence context and molecular-effect information from genomic foundation-model sequence representations and variant-effect predictors\cite{nguyen2024evo,ioannidis2016revel,jaganathan2019spliceai,cheng2023alphamissense}. Third, phenotype-conditioned biological context links the candidate gene to the patient's HPO profile through cellular, anatomical and pathway context, then computes a graph-derived representation through knowledge-guided attention\cite{velickovic2018gat,gtex2020atlas,milacic2024reactome,du2019gene2vec}. These evidence streams are fused into a patient-specific variant score.

This division of labor is data-enabled. The evidence engine is assigned the repeated, learnable part of interpretation: how rule evidence, variant-intrinsic features and phenotype-conditioned biological context should be weighted across many fetal and infant cases. The specialist modules are assigned evidence that is useful but unevenly available, computationally heavier or likely to change as external resources improve. The diagnostic agent is assigned language-facing and review-facing tasks: curation, constrained reranking, provenance audit and synthesis. In this sense, the agentic system is not only a collection of tools; it is a workflow that separates what can be learned from the in-house case distribution from what should remain callable, refreshable and inspectable.

Specialist evidence modules operate around the evidence engine. They provide additional views for candidates that warrant deeper review: phenotype-driven scores from SHEPHERD, an independent Exomiser track for candidate-pool calibration and LLM-assisted reranking, and structure-informed refinement from AlphaFold DB and Rosetta/PyRosetta-style modeling when protein modeling is meaningful\cite{smedley2015exomiser,alsentzer2025shepherd,jumper2021alphafold,varadi2024alphafold_db,fleishman2011rosettascripts,alford2017rosetta}. This layer is useful because some evidence sources are too specialized, computationally expensive or variant-class-specific to be treated as universal inputs for every candidate. DeepBD treats them as callable evidence modules whose outputs can refine ranking and support diagnostic review.

The grounded diagnostic agent sits at the end of this evidence pipeline. Its input is a ranked evidence workspace containing candidate variants, scores, source-specific evidence fields, graph traces and module outputs. The current implementation supports constrained top-k reranking, reflection-style review, evidence audit and case-level evidence synthesis within the workspace. The causal score remains anchored in the evidence engine and specialist-module outputs.

The overview also summarizes how this design is evaluated. Curated positive variants were grouped by DeepBD ranking outcome and inspected together with rule, mechanism and structure evidence patterns (Fig.~\ref{fig:overview}g), providing a compact view of how different evidence streams appear among top-ranked and missed variants. A variant-class-stratified Hit@5 comparison further shows that the benchmark covers diverse molecular consequences, including missense, frameshift, splice and truncating variants (Fig.~\ref{fig:overview}h). These panels are intended as an overview of the evidence and evaluation landscape; detailed benchmark results and ablations are reported below.

\subsection{Fetal-infant cohort}

We developed and evaluated DeepBD using an in-house cohort of 18,622 fetal and infant cases with sequencing and phenotype information (Fig.~\ref{fig:overview}c), and
evaluated ranking performance on held-out solved-case benchmarks derived from this cohort. The cohort reflects the intended clinical setting: cases are primarily detected during prenatal evaluation or early infancy and include birth-defect categories such as skeletal, cardiac, fetal growth, nuchal translucency or RASopathy-related, central nervous system, renal and developmental phenotypes. This scope is deliberate. Prenatal genetic diagnosis and infant genetic diagnosis are often presented as separate workflows, but genetic birth defects frequently move across this boundary as the same fetus is imaged, sequenced, delivered, examined and reinterpreted over time\cite{vora2023prenatal,p3egs2023diagnostic}. A fetal-to-infant cohort therefore provides a more appropriate substrate for this disease space than a purely adult rare-disease benchmark or a prenatal-only demonstration.

This paired phenotype-sequencing resource is also part of the method rather than only an evaluation set. Public genetic-disease resources contain valuable curated genes, variants, ontology terms and case reports, but they are usually fragmented across disease databases, variant archives, phenotype ontologies, literature and tool-specific annotations. They rarely provide, at scale, the exact combination needed to train a birth-defect prioritization model: fetal or early-infant phenotypes, sequencing-derived candidate variants, clinically curated causal labels and realistic non-causal competing variants from the same patient. The in-house cohort allowed DeepBD to learn the distribution of candidate variants encountered in this clinical setting before supervised optimization on solved cases. This is the empirical basis for assigning stable, repeated evidence integration to the pretrained evidence engine while leaving rapidly changing or highly specialized evidence to modular retrieval and review.

This point also clarifies why the cohort is a contribution rather than background bookkeeping. Learning-based prioritizers in Mendelian and rare-disease diagnosis have benefited from real diagnosed cases, curated pathogenic variants, simulated patients or broad clinical-genomic repositories\cite{boudellioua2019deeppvp,danzi2023maverick,mao2024aim,alsentzer2025shepherd}. Those resources are valuable, but they do not fully capture the fetal-to-infant birth-defect setting in which phenotypes are often imaging-derived, developmentally incomplete and linked to time-sensitive reproductive or neonatal decisions. The in-house cohort gives DeepBD access to the competing-variant distribution that clinicians actually face after sequencing in this setting, which is precisely the information needed to train a ranking substrate rather than relying only on fixed rules or open-ended agentic reasoning.

The solved-case benchmark used for ranking evaluation contained clinically curated positive variants. The held-out ranking split was sampled at the proband level and contained 549 proband cases with 1,307 curated positive variants. Retained candidate variants spanned 22 categories and included a broad range of predicted molecular consequences (Fig.~\ref{fig:overview}d). The distribution motivates multi-level evidence integration. Some positive variants are supported by established clinical databases or clear loss-of-function rules; others require weaker but complementary evidence from phenotype match, cellular or pathway context, sequence-based effect prediction, external prioritization tools or protein-structure analysis. The in-house dataset is not publicly releasable because it contains sensitive fetal, infant, genomic and hospital-derived clinical information, but its scale allowed evidence-model pretraining, supervised ranking development and ablation in a data regime that is difficult to obtain for genetic birth defects.

The cohort also shaped the intended use of the workflow. In a prenatal-only workflow, phenotype information is often dominated by imaging descriptors and may be missing postnatal manifestations. In a general rare-disease workflow, the phenotype may include years of longitudinal signs, laboratory results and specialist notes. The fetal and infant cases used here lie between these extremes. They include structural findings that are visible before birth, early neonatal signs that emerge after delivery and molecular results that must be interpreted before the phenotype is fully known. DeepBD therefore treats the phenotype profile as a patient-specific context rather than as a complete disease description.

\begin{figure}[!t]
\centering
\includegraphics[width=\linewidth]{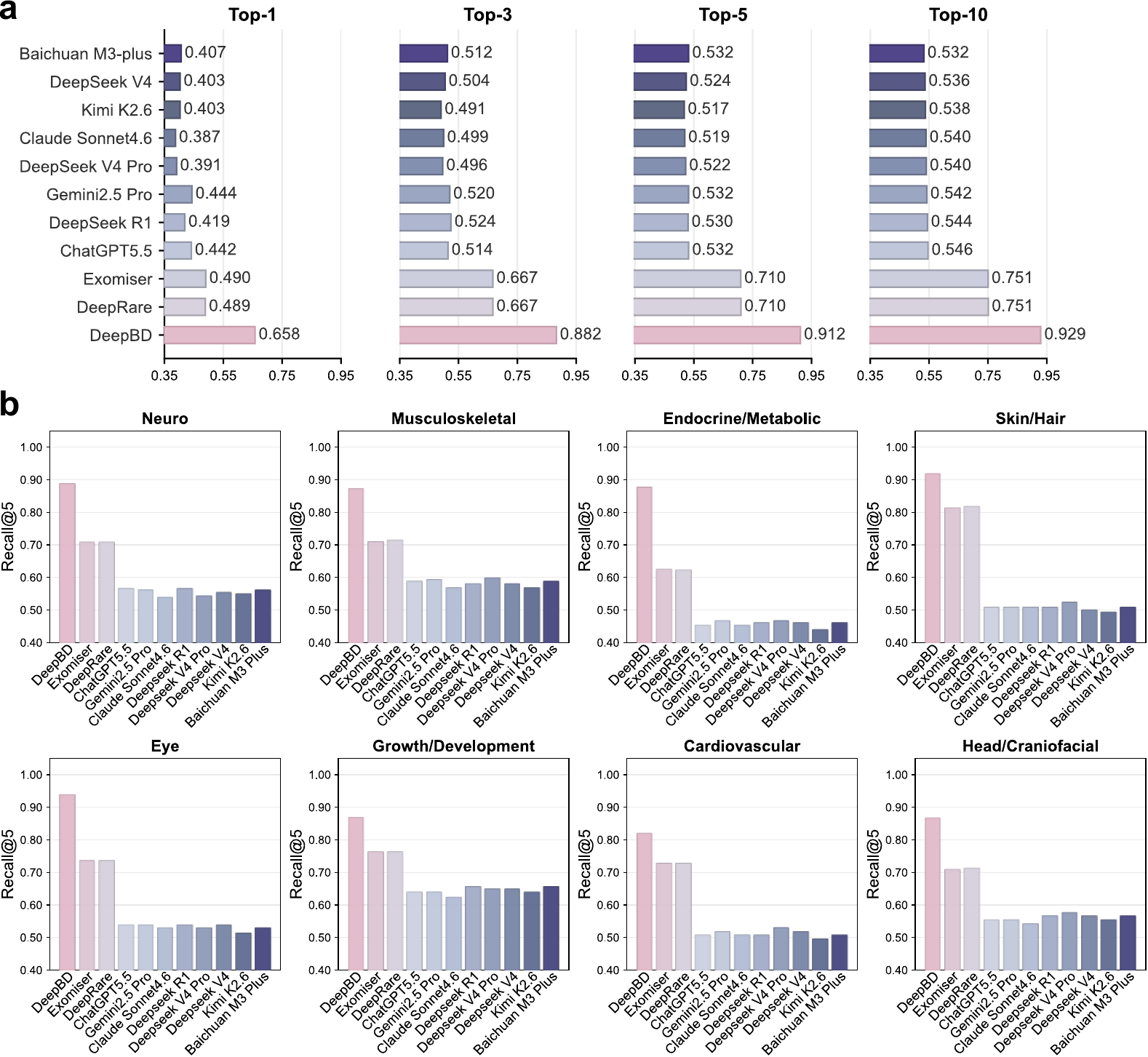}
\caption{\textbf{Benchmarking DeepBD for causal variant prioritization.}
\textbf{a,} Overall Recall@1, Recall@3, Recall@5 and Recall@10 for DeepBD, standalone Exomiser, DeepRare and prompted LLM reranking baselines. DeepBD achieved 0.658, 0.882, 0.912 and 0.929, respectively.
\textbf{b,} Recall@5 stratified by phenotype category, showing performance across major fetal and infant birth-defect presentations.}
\label{fig:benchmark}
\end{figure}

\subsection{Variant-level reranking benchmark}

The evaluation was designed to resemble the practical question faced after sequencing: among retained candidate variants for a fetal or infant case, how highly does the system rank the curated positive variant or variants? We therefore used case-level Recall@K rather than global binary pathogenicity classification. Recall@K measures whether a curated positive variant appears within the first K candidates for manual review, and is clinically meaningful because a geneticist usually examines a small number of prioritized candidates before returning to broader filtering. This variant-level endpoint is appropriate for post-sequencing interpretation, where the actionable unit is a specific variant, inheritance pattern and evidence package rather than a gene name alone.

We compared DeepBD with three baseline groups. The first group consisted of standalone Exomiser, an established phenotype-driven variant and gene prioritization tool\cite{smedley2015exomiser}. The second group consisted of prompted general and reasoning LLMs. Because general-purpose LLMs cannot reasonably process full VCF-scale candidate spaces, they were evaluated as rerankers over the Exomiser-derived top-20 candidate variants for each case; they received structured variant and phenotype information without the Exomiser rank order. The third group consisted of DeepRare, an agentic rare-disease diagnostic system\cite{zhao2026deeprare}. We report DeepRare under this configuration because it matches its native tool-use setting; however, we treat this as a reranking
baseline rather than a full reproduction of DeepRare’s end-to-end diagnostic workflow. Thus, LLM and agentic baselines were tested in a tractable candidate-reranking setting rather than as full VCF-to-diagnosis pipelines.

This setting also distinguishes ranking performance from diagnostic explanation. DeepBD's primary quantitative endpoint is variant prioritization. The diagnostic agent is included because interpretation and diagnosis require provenance-preserving synthesis, but its narrative output is not treated as independent proof of correctness. The comparison therefore isolates evidence weighting after candidate preprioritization, with diagnostic synthesis evaluated separately.

\subsection{Causal variant prioritization}

On the internal held-out solved-case ranking benchmark, DeepBD achieved Recall@1, Recall@3, Recall@5 and Recall@10 of 0.658, 0.882, 0.912 and 0.929, respectively (Fig.~\ref{fig:benchmark}a). Standalone Exomiser achieved 0.490, 0.667, 0.710 and 0.751, and DeepRare achieved 0.489, 0.667, 0.710 and 0.751 in the same Exomiser-derived candidate-reranking setting. Prompted LLM baselines showed lower overall recall, with top-1 recall ranging from 0.387 to 0.444 and top-10 recall ranging from 0.532 to 0.546. Phenotype-stratified Recall@5 showed the same direction across major categories (Fig.~\ref{fig:benchmark}b). DeepBD's largest absolute gains occurred beyond rank 1, indicating that the workflow expands the set of cases in which a curated positive variant appears within the short list used for manual review.

These results should be interpreted as variant-level reranking results, not as disease-level differential diagnosis results. This distinction is important because several baselines were originally designed for different tasks. The comparison nevertheless addresses a clinically relevant step: after conventional candidate preprioritization has produced a short list, a system must decide which variant should be reviewed first and which evidence supports that decision. The gain over standalone Exomiser indicates that DeepBD is not only inheriting the candidate set from conventional prioritization, but also changing the relative ordering through patient-specific evidence integration.

\subsection{Complementary evidence streams}

\begin{figure}[!tp]
\centering
\includegraphics[width=\linewidth]{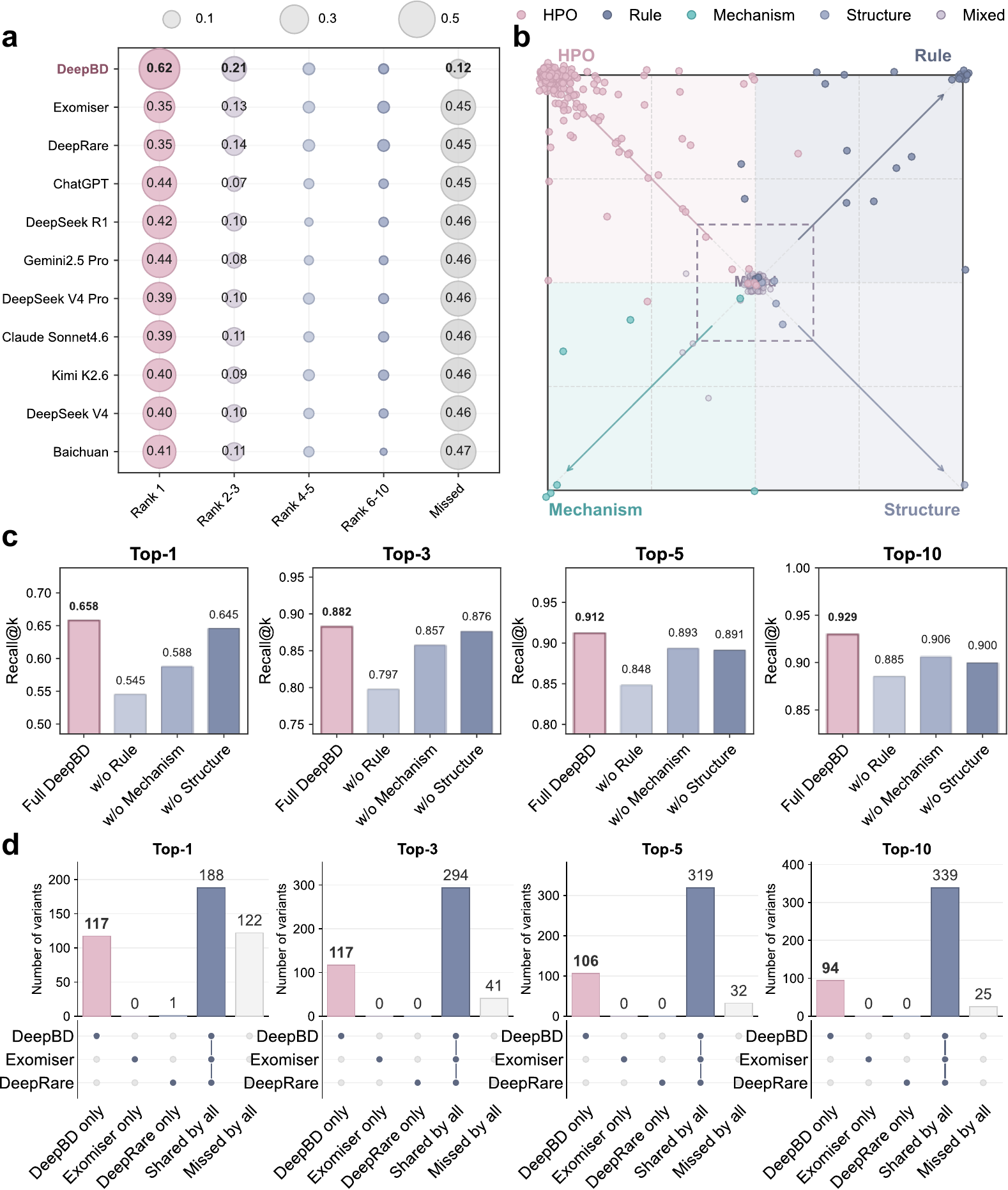}
\caption{\textbf{Evidence grounding, complementarity and ablation.}
\textbf{a,} Distribution of causal-variant ranks for DeepBD and baseline methods.
\textbf{b,} Case-level evidence landscape summarizing the relative contribution of HPO, rule, mechanism, structure and mixed evidence patterns.
\textbf{c,} Ablation analysis. Removing rule, graph-derived mechanism or structure-informed evidence decreases Recall@K, indicating that DeepBD relies on multi-level evidence integration.
\textbf{d,} Overlap analysis of cases prioritized by DeepBD, standalone Exomiser and DeepRare at different K values.}
\label{fig:evidence}
\end{figure}

Rank-distribution analysis showed that DeepBD increased the fraction of cases solved at rank 1 and reduced the missed-case fraction relative to baseline tools and LLMs (Fig.~\ref{fig:evidence}a). The evidence landscape suggested that cases are not supported by a single uniform evidence type (Fig.~\ref{fig:evidence}b). Some are dominated by HPO-driven phenotype fit, others by explicit rule evidence, mechanism context, structure-informed refinement or mixed patterns. This heterogeneity is expected in genetic birth defects, where the same clinical endpoint may arise from well-curated recurrent variants, sparse gene-disease relationships, uncertain variant effects or context-dependent phenotype matches.

Ablation experiments confirmed that the system depends on multiple evidence levels (Fig.~\ref{fig:evidence}c).  Removing explicit rule evidence reduced Recall@1 to 0.474 and Recall@10 to 0.639. Removing graph-derived mechanism evidence reduced Recall@1 to 0.494 and Recall@10 to 0.630. Removing structure-informed evidence reduced Recall@1 to 0.521 and Recall@10 to 0.741. 
These results support two conclusions.  First, explicit clinical rules remain essential; learned representations should complement population frequency, curated pathogenicity, dosage sensitivity and gene-disease validity evidence. Second, graph-derived context and structure-informed refinement contribute beyond conventional annotations, indicating that DeepBD's gain is distributed across the evidence substrate rather than driven by a single isolated module.

Overlap analysis with standalone Exomiser and DeepRare further showed that DeepBD recovered cases that were missed by both baseline ranking systems: 117 cases at top 1, 117 at top 3, 106 at top 5 and 94 at top 10 were uniquely prioritized by DeepBD among the compared methods (Fig.~\ref{fig:evidence}d). This complementarity is important because it shows that DeepBD is not merely reproducing the same candidates prioritized by existing systems. Rather, the workflow changes the relative ordering of variants in a way that produces additional top-ranked recoveries after conventional candidate preprioritization.

\subsection{Diagnosis-oriented review}

DeepBD's diagnostic agent is intentionally grounded in the evidence substrate. It receives ranked candidates and their evidence traces, organizes them into a reviewable workspace and supports diagnosis-oriented synthesis. In the current implementation, this layer includes constrained LLM-assisted top-k reranking and reflection-style review over structured candidate records, with companion evidence-analysis prompts for organizing candidate mechanisms. For each candidate, the review layer can expose population-frequency signal, consequence annotation, variant-effect predictions, gene-disease validity, ClinVar or HGMD evidence, phenotype-linked graph paths, high-attention cellular or pathway context, external tool signals and structure-informed evidence when available. This structure is designed to preserve provenance so that a clinician can inspect the sources underlying a candidate diagnosis.

This design differs from an LLM-first diagnostic workflow. In an LLM-first system, the agent often performs retrieval, reasoning and final diagnosis in one loop. Such systems can be powerful when the task is broad differential diagnosis, especially when the input is free-text clinical narrative and the desired output is a list of diseases with supporting rationale\cite{zhao2026deeprare,liu2026hygieia,meng2026berrylyzer}. For post-sequencing interpretation of genetic birth defects, the key decision is a ranked comparison among patient-specific variants. DeepBD places the agentic layer around evidence computation, where it can improve usability, evidence audit and diagnosis-oriented synthesis while the causal ranking remains benchmarkable.

A top-ranked variant is not automatically a diagnosis; it must be connected to inheritance, phenotype fit, molecular consequence and disease validity. DeepBD's output is therefore framed as a ranked evidence table plus a diagnosis-oriented hypothesis rather than as a black-box disease label. This is important for fetal and infant cases, where the same molecular result may lead to prognosis discussion, confirmatory testing, parental testing, postnatal follow-up or later reanalysis. In this study, the primary quantitative claim is variant prioritization; diagnostic report quality, factuality and clinical usefulness require separate expert evaluation.

\section{Discussion}

DeepBD addresses post-sequencing interpretation of genetic birth defects by combining a pretrained evidence engine, specialist evidence modules and a grounded agentic review layer. The key design choice is evidence allocation. Core variant prioritization is performed by a trainable model that integrates explicit clinical rules, sequence and variant-effect information and phenotype-conditioned biological context. Specialist modules then refine or contextualize selected candidates, and agentic components operate around the resulting evidence table to structure cases, coordinate evidence use, organize provenance and support diagnostic synthesis. This architecture is well matched to a high-stakes genomic workflow in which performance can be measured by ranking accuracy and explanations must remain traceable to evidence sources.

The work extends phenotype-driven variant prioritization in a birth-defect-context-aware direction. Established systems such as Phen-Gen, Exomiser, Xrare and LIRICAL showed that phenotype and genotype information should be interpreted together\cite{javed2014phengen,smedley2015exomiser,li2019xrare,robinson2020lirical}. DeepBD retains this principle but represents the candidate variant in a broader patient-specific context: the phenotype profile conditions biological context around each candidate gene, and this representation is fused with variant-intrinsic and rule-based evidence before selected candidates are refined by specialist modules. The resulting model does not claim to simulate human development. Its narrower and more defensible role is to compute phenotype-conditioned biological context for birth defect interpretation.

The work also reframes how agentic AI can be used in genomic diagnosis. LLMs and agents are attractive because they can parse clinical language, retrieve evolving external knowledge, coordinate tools and generate human-readable diagnostic narratives\cite{singhal2023medpalm,tu2025amie,zhao2026deeprare,liu2026hygieia,meng2026berrylyzer}. In variant interpretation, free-form reasoning is a fragile foundation for causal ranking unless it is constrained by structured evidence. DeepBD therefore uses a grounded agentic design: the agentic layer receives ranked candidates, evidence vectors and source-specific traces, and its role is to synthesize and communicate rather than to invent the underlying causal score. This does not diminish the value of agents; it places them at the stage where their strengths are most useful and their risks are easier to control.

The fetal and infant cohort is a practical strength. Genetic birth defects occupy a continuum that begins with prenatal imaging and may continue through neonatal or infant follow-up. A purely prenatal framing would understate the role of postnatal phenotype evolution, whereas a broad rare-disease framing would obscure the specific challenges of fetal imaging, developmental incompleteness and birth-defect categories. The 18,622-case in-house cohort provides a disease-context-rich substrate for model development, although privacy and data-use constraints limit public release. The present results therefore support the value of a birth-defect-focused benchmark, while also highlighting the need for external validation across institutions.

The layered architecture also provides a clearer path for validation than a single end-to-end diagnostic chatbot. Each evidence stream can be inspected by a domain expert, removed in ablation, refreshed when external knowledge changes and linked back to the ranked candidate that used it. This is useful in birth defect interpretation because evidence changes over time. Prenatal imaging can be reinterpreted after additional scans, neonatal findings can add new HPO terms, parental testing can clarify inheritance and newly published gene-disease relationships can change the plausibility of a candidate. A system that separates phenotype normalization, variant ranking, biological-context modeling, tool-based refinement and diagnostic review can support this iterative workflow more naturally than a one-shot prediction pipeline.

The clinical role of such a system should therefore be understood as prioritization and evidence organization. A high DeepBD rank can accelerate expert review by bringing plausible causal variants forward and by showing which evidence streams support that rank. A low rank should not exclude clinical judgment, especially for variant classes, inheritance patterns or genes that are under-represented in current resources. This division is important for responsible deployment: the model provides a measurable ranking function, and the review layer provides a structured place to examine supporting evidence, conflicting evidence and missing evidence before any diagnosis is finalized.

Several limitations remain. The cohort is single-institution or collaboration-derived and requires external validation across sequencing pipelines, hospitals and ancestry groups. The current benchmark focuses on variant prioritization in solved or curated cases; diagnostic-report quality, clinical actionability and prospective workflow impact require additional expert review. The LLM and agentic baselines were evaluated as rerankers over Exomiser-derived top-20 candidates, which is clinically tractable but distinct from full VCF-scale discovery or broad disease-level differential diagnosis. Structure-informed refinement is useful for certain missense or high-priority uncertain variants but cannot be applied uniformly across all variant classes. The agentic diagnostic layer is intentionally lightweight in the present version and should be evaluated separately from the ranking substrate as it matures. Future work will extend evaluation to external cohorts, perform prospective clinical validation and test whether provenance-preserving evidence tables improve multidisciplinary review in prenatal and neonatal genetics.

More broadly, DeepBD suggests a practical path for applying AI to genomic medicine. Rather than treating diagnosis as an unconstrained language-generation problem, the system organizes diagnosis as an evidence-to-diagnosis workflow. The evidence engine produces a quantitative, testable ordering of variants; graph and structure modules provide biological context; rule evidence anchors predictions to clinical resources; specialist tools provide additional calibration or refinement; and the agentic layer turns these outputs into a format that can be reviewed and updated. This layered design is especially relevant for genetic birth defects, where the diagnostic pathway is time-sensitive, phenotype information is incomplete and evidence often changes as the case evolves.

\bibliography{sample}

\clearpage
\section{Methods}

\subsection{Problem formulation}

For each patient case $i$, DeepBD receives a phenotype profile and a set of candidate variants,
\begin{equation}
\mathcal{C}_i = (\mathcal{P}_i, \mathcal{V}_i), \quad
\mathcal{P}_i=\{p_{i1},\ldots,p_{im}\}, \quad
\mathcal{V}_i=\{v_{i1},\ldots,v_{in_i}\}.
\end{equation}
The phenotype profile $\mathcal{P}_i$ consists of HPO terms or text-derived phenotype concepts. The candidate set $\mathcal{V}_i$ is obtained from a VCF- or TSV-derived variant table after standard annotation and filtering. The primary goal is to learn a patient-specific scoring function $S_\theta$ that assigns each candidate variant $v_{ij}$ a priority score,
\begin{equation}
s_{ij}=S_\theta(v_{ij}, \mathcal{P}_i, \mathcal{K}), \quad s_{ij}\in[0,1],
\end{equation}
where $\mathcal{K}$ denotes external biomedical knowledge sources. Candidates are sorted by $s_{ij}$ to produce a ranked list $\pi_i$. The review layer receives $\pi_i$ and evidence traces $\mathcal{E}_i$ and produces diagnosis-oriented synthesis $D_i$,
\begin{equation}
(\pi_i,\mathcal{E}_i,D_i)=\mathrm{DeepBD}(\mathcal{P}_i,\mathcal{V}_i,\mathcal{K}).
\end{equation}
The present manuscript evaluates the ranking component using solved cases with curated positive variants. Diagnostic synthesis is reported as a workflow component and requires separate expert evaluation.

\subsection{Phenotype and variant inputs}

Phenotype inputs are represented as HPO identifiers whenever structured annotations are available. When clinical descriptions are provided as text, the curation layer maps phrases to HPO concepts and preserves the original phrase, matched term and matching confidence. The normalized phenotype set is used by the graph module and retained as review evidence.

Variant inputs are derived from VCF or tabular annotation files. Each candidate variant includes chromosome, position, reference allele, alternate allele, gene symbol, transcript, HGVS notation when available, predicted consequence, zygosity, read depth, genotype quality and other sequencing fields. The system processes one case at a time. Within a case, all retained candidate variants are scored in parallel and compared only with other variants from the same patient.

\subsection{Core evidence representation}

For each candidate variant $v_{ij}$, DeepBD constructs a set of evidence representations rather than a single annotation score. The rule branch extracts an explicit evidence vector
\begin{equation}
\mathbf{x}^{\mathrm{rule}}_{ij}
=[\mathbf{x}^{\mathrm{basic}}_{ij};\mathbf{x}^{\mathrm{gene}}_{ij};\mathbf{x}^{\mathrm{clinical}}_{ij}]
\in \mathbb{R}^{d_R}.
\end{equation}
The basic component encodes population allele frequency, sequencing quality and predicted impact. Population frequencies are queried from gnomAD and related population resources\cite{karczewski2020gnomad,lek2016exac,1000genomes2015}. The gene component summarizes ClinGen gene-disease validity, dosage sensitivity, gene constraint and related gene-level priors\cite{strande2017clingen}. The clinical component summarizes ClinVar, HGMD, PubMed-derived evidence, inheritance compatibility and phenotype-conditioned disease priors\cite{landrum2016clinvar,stenson2020hgmd}. These features are designed as auditable priors for prioritization and should not be interpreted as a replacement for manual ACMG/AMP classification\cite{richards2015acmg}.

The variant branch constructs an intrinsic variant representation from sequence and functional evidence. Let $\mathbf{e}^{\mathrm{seq}}_{ij}$ denote a sequence-context representation around the variant, initialized or derived from a genomic foundation-model encoder such as Evo\cite{nguyen2024evo}. Let $\mathbf{x}^{\mathrm{func}}_{ij}$ denote functional predictors and annotation scores, including REVEL, PrimateAI, SpliceAI, AlphaMissense and impact-derived features\cite{ioannidis2016revel,sundaram2018primateai,jaganathan2019spliceai,cheng2023alphamissense}. These inputs are concatenated and projected to a dense vector,
\begin{equation}
\mathbf{h}^{V}_{ij}
=\phi_V\left(\left[\mathbf{e}^{\mathrm{seq}}_{ij};\mathbf{x}^{\mathrm{func}}_{ij}\right]\right)
\in \mathbb{R}^{d_V},
\end{equation}
where $\phi_V$ is a neural projection with normalization and nonlinear activation.

\subsection{Graph construction}

For each candidate gene $g_{ij}$, DeepBD builds a heterogeneous subgraph
\begin{equation}
\mathcal{G}_{ij}=(\mathcal{N}_{ij},\mathcal{R}_{ij})
\end{equation}
containing the candidate gene, patient HPO terms, cellular or anatomical context nodes and pathway nodes. Gene nodes are initialized using gene embeddings and gene-level metadata\cite{du2019gene2vec}. Phenotype nodes are initialized using ontology or biomedical text embeddings based on HPO resources\cite{robinson2008hpo,kohler2021hpo}. Cellular, anatomical and pathway nodes are derived from expression, anatomical and pathway resources such as GTEx and Reactome\cite{gtex2020atlas,milacic2024reactome}. Edges encode phenotype-to-context, context-to-gene and pathway-to-gene relations with prior weights $\rho_{uv}$ derived from curated databases, ontology distance or precomputed association scores. This graph is used to compute phenotype-conditioned biological context; it is not intended to be a literal simulator of embryonic development.

\subsection{Graph attention}

The graph encoder uses knowledge-guided graph attention. In the current implementation, scalar edge priors, optional edge-feature vectors and gene-conditioned trust coefficients jointly modulate message passing. For an edge from node $u$ to node $v$, a simplified attention logit can be written as
\begin{equation}
e_{uv}^{(r)}
=\mathrm{LeakyReLU}\left(\mathbf{a}^{\top}
[\mathbf{W}\mathbf{z}_u \Vert \mathbf{W}\mathbf{z}_v]\right)
+ \psi_r(\mathbf{q}_{uv})+\lambda_i^{(r)}\log(\rho_{uv}+\epsilon),
\end{equation}
where $\mathbf{z}_u$ and $\mathbf{z}_v$ are node embeddings, $r$ denotes relation type, $\rho_{uv}$ is a scalar edge prior, $\mathbf{q}_{uv}$ denotes optional edge features such as HPO--gene or pathway--gene association statistics, $\psi_r$ is an edge-feature projection and $\epsilon$ prevents numerical instability. The coefficient $\lambda_i^{(r)}$ is generated for each candidate-gene context from the gene embedding and gene-level reliability features such as literature count, ClinVar record count and pathway-network degree. The normalized attention coefficient is
\begin{equation}
\alpha_{uv}^{(r)}=\frac{\exp(e_{uv}^{(r)})}{\sum_{u'\in\mathcal{N}(v)}\exp(e_{u'v}^{(r)})}.
\end{equation}
The trust coefficient can be expressed as
\begin{equation}
\lambda_i^{(r)}=\lambda_{\min}+\gamma\cdot
\sigma\left(f_{\tau}^{(r)}([\mathbf{z}_{g_{ij}};\mathbf{d}_{g_{ij}}])\right),
\end{equation}
where $\mathbf{d}_{g_{ij}}$ denotes gene-level evidence-density features. When latent phenotype-to-cell routing is enabled, patient HPO embeddings are pooled into a phenotype-set query and routed to cellular context nodes, with optional prior bias. The graph output is a phenotype-conditioned gene-context vector,
\begin{equation}
\mathbf{h}^{G}_{ij}=\phi_G(\mathcal{G}_{ij},\mathcal{P}_i)\in \mathbb{R}^{d_G}.
\end{equation}
High-attention phenotype, cellular, anatomical and pathway nodes are retained as provenance-preserving evidence traces when available.

\subsection{Evidence fusion}

The evidence engine fuses variant-intrinsic evidence, phenotype-conditioned graph evidence and rule evidence. A generic interaction-aware fusion can be written as
\begin{equation}
\mathbf{m}_{ij}
=\phi_M\left(
[\mathrm{LN}(\mathbf{h}^{V}_{ij});
\mathrm{LN}(\mathbf{h}^{G}_{ij});
\mathrm{LN}(\mathbf{h}^{V}_{ij})\odot \mathrm{LN}(\mathbf{h}^{G}_{ij})]
\right).
\end{equation}
This vector should be interpreted as a learned representation of variant effect in patient phenotype context. It is a computational link between variant-intrinsic evidence and birth-defect context.

The rule vector is projected to a dense representation,
\begin{equation}
\mathbf{r}_{ij}=\phi_R(\mathbf{x}^{\mathrm{rule}}_{ij})\in\mathbb{R}^{d'_R}.
\end{equation}
DeepBD then combines mechanism and rule representations through feature-wise gating and cross-evidence interaction. Let
\begin{equation}
\mathbf{z}_{ij}=[\mathbf{m}_{ij};\mathbf{r}_{ij}].
\end{equation}
The gated representation is
\begin{equation}
\tilde{\mathbf{z}}_{ij}=
\sigma(\mathbf{W}_{g}\mathbf{z}_{ij}+\mathbf{b}_{g})\odot \mathbf{z}_{ij}.
\end{equation}
The cross-interaction representation is
\begin{equation}
\mathbf{c}_{ij}=
(\mathbf{W}_{m}\mathbf{m}_{ij})\odot(\mathbf{W}_{r}\mathbf{r}_{ij}).
\end{equation}
The final priority score is
\begin{equation}
s_{ij}=\sigma\left(\mathbf{w}^{\top}
\phi_S([\tilde{\mathbf{z}}_{ij};\mathbf{c}_{ij}])+b\right),
\end{equation}
where $\phi_S$ is a residual multilayer perceptron with batch normalization. Candidate variants are ranked by $s_{ij}$ within each patient.

\subsection{Specialist modules}

Specialist evidence modules provide additional evidence for selected candidates or top-ranked subsets. Each module $\mathcal{M}_{\ell}$ receives a candidate variant, phenotype profile and module-specific resources,
\begin{equation}
\mathbf{u}_{ij}^{(\ell)}=\mathcal{M}_{\ell}(v_{ij},\mathcal{P}_i,\mathcal{K}_{\ell}).
\end{equation}
Modules include phenotype-driven SHEPHERD scores\cite{alsentzer2025shepherd}, an independent Exomiser track for candidate-pool calibration and LLM-assisted reranking\cite{smedley2015exomiser}, and structure-informed refinement for variants where protein modeling is meaningful. A generic calibrated score can be written as
\begin{equation}
\bar{s}_{ij}=\mathcal{A}_{\omega}\left(s_{ij}, \mathbf{u}_{ij}^{(1)}, \ldots, \mathbf{u}_{ij}^{(L)}\right),
\end{equation}
where $\mathcal{A}_{\omega}$ denotes a calibrated or rule-constrained aggregation function. In the present implementation, module outputs are used as refinement and review evidence through weighted score adjustment, candidate-pool construction or LLM review prompts rather than as unconstrained agent-generated scores.

For candidate missense variants or high-priority uncertain variants, DeepBD can retrieve wild-type protein structures from AlphaFold DB\cite{jumper2021alphafold,varadi2024alphafold_db}. Local structural features are computed using Rosetta/PyRosetta-style modeling\cite{fleishman2011rosettascripts,alford2017rosetta}. The structure feature vector can include local confidence, residue environment, steric or energetic change, structural deviation and physicochemical property change,
\begin{equation}
\mathbf{x}^{\mathrm{struct}}_{ij}
=[\mathrm{pLDDT},\Delta\Delta G,\Delta\mathrm{RMSD},\Delta q,\Delta \mathrm{hydrophobicity},\ldots].
\end{equation}
These values are used as refinement and reporting evidence and are included in ablation analysis.

\subsection{Evidence pretraining}

Before supervised ranking optimization, DeepBD initializes the evidence encoder using self-supervised contrastive pretraining on in-house fetal and infant variant tables without using causal-variant labels. For each candidate variant record, two independently masked views are generated by randomly masking parts of the structured variant and evidence fields. The same evidence model encodes the two views into embeddings
\begin{equation}
\mathbf{u}_{ij}^{(a)} = E_{\theta}(\tilde{v}_{ij}^{(a)}, \mathcal{P}_i, \mathcal{K}), \quad
\mathbf{u}_{ij}^{(b)} = E_{\theta}(\tilde{v}_{ij}^{(b)}, \mathcal{P}_i, \mathcal{K}),
\end{equation}
where $\tilde{v}_{ij}^{(a)}$ and $\tilde{v}_{ij}^{(b)}$ denote two masked views of the same candidate variant. The contrastive objective encourages embeddings from the same variant record to be close and embeddings from different records in the minibatch to be separated,
\begin{equation}
\mathcal{L}_{\mathrm{pre}}
=-\frac{1}{B}\sum_{j=1}^{B}
\log
\frac{\exp(\mathrm{sim}(\mathbf{u}_{j}^{(a)},\mathbf{u}_{j}^{(b)})/\tau)}
{\sum_{k=1}^{B}\exp(\mathrm{sim}(\mathbf{u}_{j}^{(a)},\mathbf{u}_{k}^{(b)})/\tau)}.
\end{equation}
Here $\mathrm{sim}(\cdot,\cdot)$ is cosine similarity after normalization and $\tau$ is a temperature parameter. The pretrained parameters are then used to initialize the evidence model for supervised causal-variant ranking.

\subsection{Training objective}

For each solved training case, curated positive variants are assigned high labels and other retained candidates are treated as within-case background or lower-confidence candidates. In the current implementation, variants with $y_{ij}\geq 0.8$ are treated as positives for the main ranking constraints, and variants with low labels are used as background comparators. We denote the positive set as $\mathcal{V}_i^{+}$ and the low-label background set as $\mathcal{B}_i$. DeepBD is optimized as a patient-level ranking model. The implementation uses a composite ranking objective,
\begin{equation}
\mathcal{L}_{i}
=\lambda_K\mathcal{L}_{\mathrm{topK}}
+\lambda_1\mathcal{L}_{\mathrm{top1}}
+\lambda_L\mathcal{L}_{\mathrm{list}}
+\lambda_B\mathcal{L}_{\mathrm{BCE}}
+\lambda_A\mathcal{L}_{\mathrm{aux}},
\end{equation}
with coefficients selected by the training configuration. The top-$K$ component encourages every curated positive variant to score above the empirical $K$-th candidate in the same patient,
\begin{equation}
\mathcal{L}_{\mathrm{topK}}
=\frac{1}{|\mathcal{V}_i^{+}|}
\sum_{v_{ij}\in \mathcal{V}_i^{+}}
\left[s_{i,(K)}-s_{ij}+m_K\right]_+,
\end{equation}
where $s_{i,(K)}$ is the $K$-th largest score among candidates for case $i$ and $m_K$ is a margin. A top-1 hinge term similarly separates positives from high-scoring background candidates,
\begin{equation}
\mathcal{L}_{\mathrm{top1}}
=\left[\max_{v_{ij}\in\mathcal{B}_i}s_{ij}
-\min_{v_{ij}\in\mathcal{V}_i^{+}}s_{ij}
+m_1\right]_+ .
\end{equation}
The listwise component is used as a softer patient-level ordering signal. The implementation first maps bounded ranking scores to logits,
\begin{equation}
\ell_{ij}=\log\frac{\mathrm{clip}(s_{ij},\epsilon,1-\epsilon)}
{1-\mathrm{clip}(s_{ij},\epsilon,1-\epsilon)} ,
\end{equation}
and defines a label-weighted target distribution
\begin{equation}
q_{ij}=\frac{\exp(\beta y_{ij})}{\sum_{k=1}^{n_i}\exp(\beta y_{ik})},
\end{equation}
where $\beta$ controls the sharpness of the target distribution. The predicted distribution is
\begin{equation}
p_{ij}=\frac{\exp(\ell_{ij})}{\sum_{k=1}^{n_i}\exp(\ell_{ik})}.
\end{equation}
The corresponding loss is
\begin{equation}
\mathcal{L}_{\mathrm{list}}
=-\sum_i\sum_{j=1}^{n_i}q_{ij}\log p_{ij}.
\end{equation}
The binary term is implemented as a focal pathogenicity loss over the ranking score, and auxiliary terms regularize trust coefficients or supervise optional discovery, pathogenicity or mixture-of-experts heads during selected training stages. Ablation settings are used to test whether specific evidence streams contribute to the final ranking.

\subsection{Agentic review}

The review layer receives ranked variants and provenance-preserving evidence fields. In the current implementation, it exposes constrained LLM-assisted reranking over top-k candidates, optional dual-track review of model and Exomiser candidate pools, reflection-style quality control and structured evidence prompts for diagnosis-oriented synthesis. For each high-priority candidate, the layer can assemble population frequency, variant-effect predictions, gene validity, clinical database entries, phenotype/cellular/pathway traces, external tool signals and structural evidence into a case-level diagnostic summary. The intended output is
\begin{equation}
D_i=\mathrm{AgentReview}(\pi_i,\mathcal{E}_i,\mathcal{K}),
\end{equation}
where $D_i$ contains candidate diagnosis, supporting evidence, conflicting evidence and suggested manual-review points. In this study, the primary quantitative claim is variant prioritization; diagnostic-report quality will require separate expert evaluation.

\subsection{Evaluation metrics}

We compared DeepBD with standalone Exomiser, DeepRare and prompted LLM-based reranking systems on the same solved-case benchmark. Standalone Exomiser was evaluated as a conventional phenotype-driven variant-prioritization tool. For prompted LLM baselines, Exomiser-derived top-20 variants were used as a standardized candidate set, and the candidate order was removed before prompting. DeepRare was evaluated on the same top-20 candidates but was allowed to access the Exomiser-derived ordering because Exomiser is part of its native tool-use setting. The main metric was case-level Recall@K,
\begin{equation}
\mathrm{Recall@K}=\frac{1}{N}\sum_{i=1}^{N}
\mathbb{I}\left[\min_{v\in \mathcal{V}_i^{+}}\mathrm{rank}_i(v)\le K\right],
\end{equation}
where $\mathcal{V}_i^{+}$ is the set of curated positive variants for case $i$. We report Recall@1, Recall@3, Recall@5 and Recall@10. Ablation experiments removed rule evidence, graph-derived mechanism evidence or structure-informed evidence. Overlap analyses counted cases uniquely recovered by DeepBD or shared with baseline methods at each K.

\end{document}